\documentclass[twocolumn,nopacs,preprintnumbers]{revtex4}
\usepackage{graphicx}
\usepackage{bm}
\usepackage{color}
\usepackage[normalem]{ulem}

\usepackage{physics,dsfont}
\usepackage{subfigure}

\graphicspath{%
    {converted_graphics/}
    {/}
}

\begin{document}

\title{Time scaling and quantum speed limit in non-Hermitian Hamiltonians}

\author{F. Impens$^{1}$, F. M. D'Angelis$^{1}$, F. A. Pinheiro$^{1}$ and D. Gu\'ery-Odelin$^{2}$}

\affiliation{$^1$ Instituto de F\'{i}sica, Universidade Federal do Rio de Janeiro,  Rio de Janeiro, RJ 21941-972, Brazil
\\
$^2$ Laboratoire Collisions, Agr\'egats, R\'eactivit\'e, IRSAMC, Universit\'e de Toulouse, CNRS, UPS, France 
}

\begin{abstract}
We report on a time scaling technique to enhance the performances of quantum protocols in non-Hermitian systems. The considered time scaling involves no extra-couplings and yields a significant enhancement of the quantum fidelity for a comparable amount of resources.  We discuss the application of this technique to quantum state transfers in 2 and 3-level open quantum systems. We derive the quantum speed limit in a system governed by a non-Hermitian Hamiltonian. Interestingly, we show that, with an appropriate driving, the time-scaling technique preserves the optimality of the quantum speed with respect to the quantum speed limit while reducing significantly the damping of the quantum state norm.
\end{abstract}

\maketitle

Fast quantum control protocols have a promising future in quantum platforms as they mitigate the deleterious effects of disorder or/and dissipation. Introduced about a decade ago, shortcuts-to-adiabaticity (STA)~\cite{RMPdavid} have already a strong track record of improving quantum procedures in a wide variety of quantum platforms including ultra-cold atom setups~\cite{Schaff10,Martinez13,Amri19,Impens20}, NMR~\cite{Zhou20} and solid-state~\cite{Zhou17} systems, superconducting qubits~\cite{Wang19} and topological spin chains~\cite{FMD2020,Theocharis21}. There are several well-established methods to build STA protocols, such as the optimal control~\cite{Chasseur15}, the counterdiabatic driving~\cite{Rice03,Rice05,Rice08,Berry09}, the use of Lewis-Riesenfeld invariants~\cite{Lewis69,Chen12,Ruschhaupt12,Impens17}, or the use of properly scaled dynamical variables~\cite{Deffner14}, to name a few. Those different methods provide different strategies to hamper, compensate or mitigate the effects of non-adiabatic transitions.

As a matter of fact, STA protocols require more resources than adiabatic methods and may involve a larger number of dynamical couplings.  For instance, the Fast-Forward technique as originally introduced by Masuda~\cite{Masuda08,Masuda14,Takahashi14,Zhu21}, introduces extra couplings to be regularized in the limit of strong acceleration~\cite{Setiawan17}. The same conclusion holds for most counterdiabatic protocols. However, the dynamical control of additional interactions may constitute a limit for their practical implementation. 

In the presence of dissipation, the concept of adiabaticity breaks down. The question is rather how to mitigate the effect of dissipation for a given protocol duration, and to approach the ultimate limit provided by the quantum speed limit~(QSL)~\cite{QSL45,QSL73,QSL90,Vaidmann92,QSL03,QSL05,QSL09,QSL12,QSLOpen13a,QSLOpen13b} in non-Hermitian systems. In this article, we investigate a time scaling method for finite Hilbert spaces that tackles those two issues: it does not introduce extra couplings, enables a minimization of the resources and provides a strategy to keep the quantum speed optimality while reducing the deleterious effect of the dissipation on the state norm. 

As a starting point, we consider a given quantum trajectory $| \psi_0(t) \rangle$ solution of the time-dependent Schr\"odinger equation for the Hamiltonian $\hat{H}(t)$:
\begin{equation}
i \hbar \frac{\partial | \psi_0(t) \rangle}{ \partial t} = H(t) | \psi_0(t) \rangle,
\end{equation}
where the time-dependence of the Hamiltonian is encapsulated in a set of parameters: $\hat{H}(t)=\hat{H}[\lambda_1(t),...,\lambda_N(t)]$. The quantum trajectory $| \psi_0( \Lambda(t) ) \rangle$ is then a solution of the time-dependent Schr\"odinger equation for the rescaled Hamiltonian  $\hat{H}_{\Lambda}(t)$:
\begin{equation}
\label{eq:FFscaling}
 \hat{H}_{\Lambda}(t) = \dot{\Lambda} \hat{H}[\lambda_1(\Lambda(t)),...,\lambda_N(\Lambda(t))] \, .
 \end{equation}
 where $\Lambda(t)$ is assumed to be a monotone, differentiable function such that $\Lambda(0)=0$ and $\dot{\Lambda}(t)\geq 0$ at any time.  
 
The Hamiltonian~\eqref{eq:FFscaling} simply provides a time rescaling of the original solution. If $T$ denotes the final time at which the systems reaches the desired  quantum state target under the driving $\hat{H}(t)$, the evolution under the rescaled Hamiltonian $\hat{H}_{\Lambda}(t)$ reaches the very same target in a time $T_{\Lambda}= \Lambda^{-1}(T)$, that can be much shorter. As a result, time scaling provides a priori the simplest way to realize a shortcut to adiabaticity protocol. 

The time-scaling method provides an enhancement of the protocol performance while maintaining the original quantum trajectory.  In the following, we explain how to design the time scaling $\Lambda(t)$ in a wide variety of contexts. To work out quantitatively a strategy that minimizes the effect of dissipation, we define in  Sec.~\ref{sec:dissip} a ``quasi-unitary'' driving that ensures a constant damping rate during the whole parametrized evolution. This systematic approach provides a clear improvement over the original driving and is illustrated in 2 and 3-level systems. This strategy can be applied jointly with geometric corrections on the driving field mitigating the effects of dissipation~\cite{Impens19}. We explain how a suitable choice for the time-scaling function enables one to minimize the energetic cost of STA protocols for both closed and open systems while achieving the same quantum fidelity. In Sec.~\ref{sec:QSL}, we discuss the relation between the time-scaling transform and the quantum speed limit. Generalizations of the QSL to open systems have been obtained within the density matrix formalism~\cite{QSLOpen13a}, and in connection with the concept of Fisher information~\cite{QSLOpen13b}. Here, we put forward a simple derivation of the QSL for quantum systems driven by non-Hermitian Hamiltonians in the spirit of the Vaidmann bound~\cite{Vaidmann92}. We show that the time-scaling transform preserves the ratio of the quantum speed to the QSL in 2 and 3-level dissipative systems with appropriate corrections to the quantum driving.

 \section{Time scaling for dissipative 2 and 3-level systems}
 \label{sec:dissip}
 
We discuss here the application of the time scaling method to open quantum systems described by non-Hermitian Hamiltonian~\cite{Plenio98,Moiseyev98}.  First, we address the commonly-called FAQUAD (Fast quasiadiabatic) protocol ~\cite{FAQUAD1,FAQUAD2} for a dissipative two-level system~\cite{CohenBookPhotonAtoms}. We then investigate the application of time scaling to the STIRAP (Stimulated Raman adiabatic passage) protocol in a 3-level systems.

\subsection{FAQUAD driving in a 2-level dissipative system}

The FAQUAD protocol has been originally proposed for dissipationless quantum systems to perform a state to state transformation as quickly as possible while remaining as adiabatic as possible at all times. Let's remind the main features of this protocol for a two-level quantum system described by the control Hamiltonian:
 \begin{equation}
 \label{eq:2levelcontrolHamiltonian}
 \hat{H}_0(t)= 
  \frac {\hbar} {2} \left( \begin{array} {cc}   \delta(t) &   \Omega(t)    \\ 
\Omega^*(t) & -\delta(t) 
\end{array}  \right)
\end{equation}
expressed here in the decoupled $\{| e \rangle, | g \rangle \}$ basis. Its instantaneous eigenvalues are $E_{\pm}(t)= \pm \hbar \sqrt{\delta(t)^2+\Omega(t)^2}$. They are associated to the instantaneous eigenvectors of $ \hat{H}_0(t)$
\begin{equation}
| \phi_+(\theta ) \rangle =  \left( \begin{array} {c}   \cos (\frac{\theta}{2})   \\ 
 \sin (\frac{\theta}{2}) 
\end{array}  \right),  \; | \phi_-(\theta ) \rangle =  \left( \begin{array} {c}   \sin (\frac{\theta}{2})   \\ 
 - \cos (\frac{\theta}{2}) 
\end{array}  \right), 
\end{equation}
where $\theta(t)=\acos[\delta(t)/\sqrt{\delta(t)^2+\Omega^2(t)}]$. In the FAQUAD approach, the time evolution of $\theta(t)$ is obtained by keeping constant the adiabatic criterium (equal to a constant $c$). One finds 
 \begin{equation}
 \label{eq:FAQUADadiabaticCondition}
 \dot{\theta} = \frac {c} {\hbar} \frac {|E_+(t)-E_-(t)|} {|\langle \phi_+(\theta)| \partial_{\theta} \phi_-(\theta) \rangle |}
 \end{equation}
where $0< c <1$.  The $c \rightarrow 0$ limit is nothing but the adiabatic limit. In this protocol, non-adiabatic transitions are controlled through the ``driving speed''.

 In view of the implementation of the time-scaling method, it is worth noticing that the equality  (\ref{eq:FAQUADadiabaticCondition}) remains unchanged by the following scaling: $t \to \Lambda(t)$, $\Omega(t) \to \dot{\Lambda}(t) \Omega(\Lambda(t))$ and $\delta(t) \to \dot{\Lambda}(t) \delta(\Lambda(t))$. As a result, the scaling function $\Lambda(t)$ can be engineered to fulfill extra requirements. As an example, we propose hereafter to set this scaling function by defining the acceptable dissipation rate for the desired transformation.

We model the dissipation for this two level problem with the non-Hermitian Hamiltonian $\hat{H}=\hat{H}_0 - i \hbar\hat{\gamma}$ with $\hat{\gamma}= \gamma_e | e \rangle \langle e |+\gamma_g | g \rangle \langle g |$, where $| e \rangle$ refers to the excited state and $| g \rangle$ to the ground state. In the following, we consider an original FAQUAD passage based on the quasi-adiabatic evolution of the eigenvector $| \phi_+(\theta) \rangle$.   
As a result of the dissipation, the norm $\mathcal{N}$ of the quantum state decreases as a function of time. For a quasi-adiabatic evolution one gets
\begin{equation}
\label{eq:Schrodingerdamping}
\frac{1}{ { \mathcal{N}}}\frac {d \mathcal{N}} {d t} = -   \langle \phi_+(\theta(t)) | \hat{\gamma} | \phi_+(\theta(t)) \rangle.
\end{equation}F
The instantaneous damping rate thus depends only on the parameter $\theta$ and on the dissipation operator $\hat{\gamma}$. Taking advantage of the extra freedom provided by the time scaling function, we can also impose a fixed ``geometric'' damping rate along the trajectory:
\begin{equation}
\frac{1}{ { \mathcal{N}_\Lambda}}\frac {d \mathcal{N}_\Lambda} {d \theta} = -c'.
\label{requir1}
\end{equation}
As a result, $\mathcal{N}_\Lambda(T) = \mathcal{N}_\Lambda(0) \exp [- \pi c'  ]$.  The condition (\ref{requir1})  
prescribes a driving speed proportional to the instantaneous dissipation rate $\dot{\theta}^{\Lambda}(t)=   \langle \phi_+(\theta^{\Lambda}(t)) | \hat{\gamma} | \phi_+(\theta^{\Lambda}(t)) \rangle /c'$. It encodes mathematically the intuitive idea according to which one should increase the driving speed in region of strong dissipation. Interestingly, the driving speed depends here only on geometric features of the trajectory $| \phi_+(\theta) \rangle$, i.e. on the orientation of the associated Bloch vector, and not on its norm. Finally, the time-scaling $\Lambda(t)$ connecting the prescribed driving speed $\dot{\theta}^{\Lambda}(t)$ to the original driving $\theta(t)$ is obtained by taking the ratio of Eqs.~(\ref{eq:Schrodingerdamping}) and (\ref{requir1}):
\begin{equation}
  \label{eq:quasiunitarycondition}
 \dot{\Lambda} =   \frac {1  } { c' \dot{\theta}(\Lambda)}   \langle \phi_+  ( \theta( \Lambda  )) | \hat{\gamma} | \phi_+(\theta(\Lambda ) \rangle
    \end{equation}
The coefficient $c'$ is fixed self-consistently by the total duration of the time scaling. 
  
As a concrete example, we consider a time-rescaled FAQUAD protocol keeping the same protocol duration,  $\Lambda(T)=T$, and with a constant Rabi frequency $\Omega(t)=\Omega_0$. The single control parameter is therefore the time-dependent detuning $\delta(t)$~\cite{FAQUAD1,FAQUAD2}. Equation~\eqref{eq:FAQUADadiabaticCondition} yields $\cos \theta(t)- \cos \theta_0=  - 4  c \Omega_0 t.$  To ensure a high fidelity transfer, the angle $\theta(t)$ must fulfill the boundary conditions $\theta_0 \simeq 0$ and $\theta_T \simeq \pi$. We choose $\cos \theta_0 = 1-\epsilon$ and  $\cos \theta_T=-1+\epsilon$ with $0 <\epsilon \ll 1$ as a null parameter $\epsilon=0$ generates unrealistic infinite detuning at the time boundaries \cite{FAQUAD1,FAQUAD2}. The adiabaticity constant reads $c=(1- \epsilon)/(2 \Omega_0 T)$. One then readily finds the angle $\theta(t) = {\rm arccos}[f_{\epsilon}(t)]$ with $f_{\epsilon}(t)=(1-\epsilon)(1-2t/T)$, and the corresponding detuning $\delta(t)= \Omega / \tan \theta(t)$. We find $\langle \phi_+(\theta(t)) | \hat{\gamma} | \phi_+(\theta(t)) \rangle = \frac 1 2 (\gamma_e+\gamma_g) + \frac 1 2 (\gamma_e-\gamma_g) f_{\epsilon}(t) $ and $\dot\theta(t)=2 (1-\epsilon) / [T \sqrt{1-f_{\epsilon}(t)^2}]$. The scaling $\Lambda(t)$ is subsequently obtained by solving~\eqref{eq:quasiunitarycondition}. 

\begin{figure}[htpb]
\includegraphics[width= 0.45\textwidth]{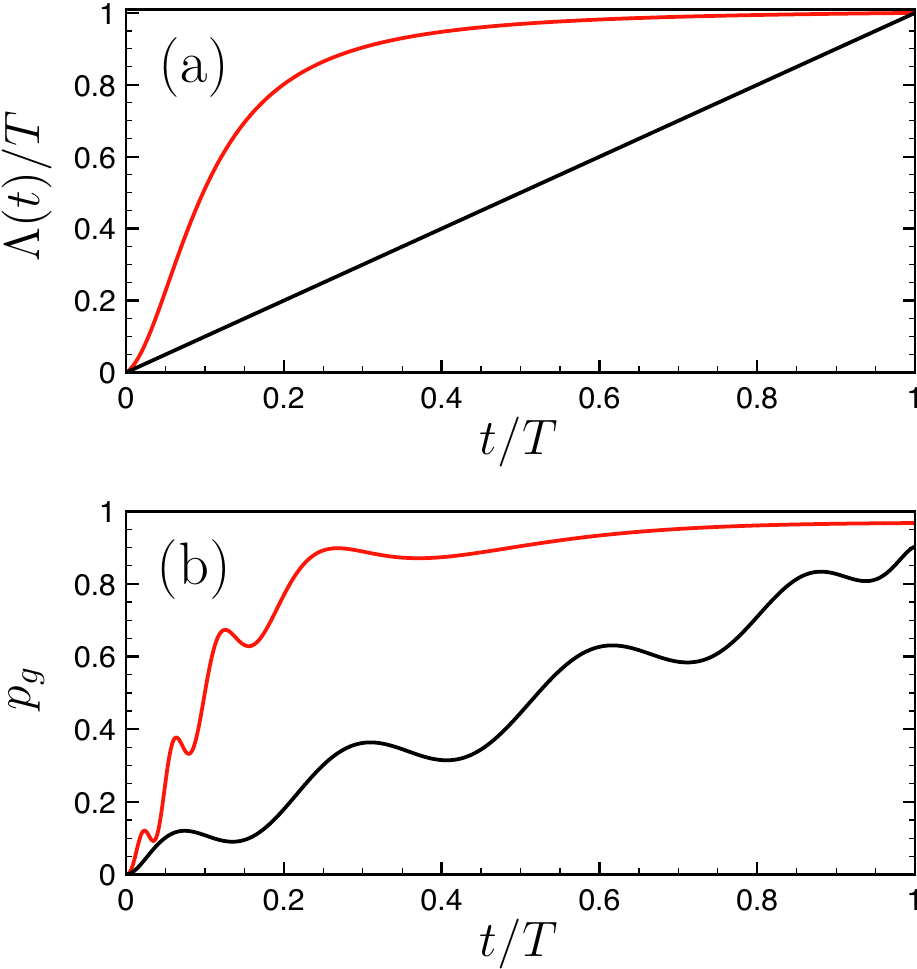}
    \caption{Time-scaled FAQUAD protocol for a 2-level system:  (a): Time-scaling function $\Lambda(t)/T$ (red solid line) as a function of the renormalized time $t/T$. The black solid line corresponds to the original protocol ($\Lambda(t)=t$).
    (b) Time-dependent occupation probability of the ground state $| g \rangle$ for the original (black solid line) FAQUAD protocol and for its time-scaled version (red solid line) as a function of the renormalized time $t/T$. Parameters: $\Omega_0 T=10$, $\epsilon=0.01$, $\gamma_{e} T = 0.1$ and  $\gamma_{g }=0.01\gamma_{e}$ .} 
	\label{fig:FAQUAD}
\end{figure}

In Fig.~\ref{fig:FAQUAD}, we summarize the results of the original FAQUAD and of its time-rescaled version for a specific example in the presence of dissipation. We have chosen the parameters $\Omega_0 T= 10$ and $\epsilon=0.01$, yielding an adiabaticity constant $c \simeq 0.05$. We use $\gamma_e T =0.1$ and $\gamma_g  =0.01 \gamma_e$. For our parameters, the condition $\Lambda(T)=T$ dictates the value of the constant $c'\simeq 5.3 \times 10^{-3}$.  The final purity $p=p_g/(p_g+p_e)$, defined as the fraction of the target state population, is $p\simeq0.998$ for both protocols. However, our time-rescaled FAQUAD protocol yields a norm reduction $\mathcal{N}_{\Lambda} \simeq 0.97$ to be compared to $\mathcal{N} \simeq 0.90$ for the initial protocol. The time-scaling thus significantly enhances the performance of the FAQUAD driving in the presence of dissipation, by keeping a high purity while reducing the norm reduction rate by at least a factor of $3$ in this specific case.

 \subsection{Time scaling in a STIRAP transfer} 
  \label{subsec:TimeScalingSTIRAP}
 
 \subsubsection{The dissipationless STIRAP solution}
 
 In this section, we investigate the interest of time scaling for an accelerated population transfer in a dissipative 3-level system. More precisely, we consider a 3-level system in a $\Lambda$-configuration. In the absence of dissipation, the quantum state, $| \psi (t) \rangle = C_1(t) |1\rangle + C_2(t) |2\rangle + C_3(t) |3\rangle$, obeys the Schr\"odinger equation associated to the control Hamiltonian 
\begin{equation}
\label{eq:STIRAPControlHamiltonian}
\hat{H}_0(t)= \frac {\hbar} {2} \left( \begin{array} {ccc}   0 &   \Omega_p^0(t) & 0   \\ 
\Omega_p^0(t) & 0  &   \Omega_s^0(t) \\
0 & \Omega_s^0(t) & 0 \\
\end{array}  \right).
\end{equation}
The transfer of the population from the ground state $|1\rangle$ to the ground state $|3\rangle$ can be realized by following adiabatically the dark state, which amounts to applying Stokes ($\Omega_s$) and pump ($\Omega_p$) field pulses with a slight delay whilst maintaining a significant temporal overlap between the two pulses \cite{ReviewSTIRAP1,ReviewSTIRAP2}.  Using an invariant-based inverse engineering technique, such a transfer can be accelerated at the expense of a transient population in the excited state $|2\rangle$~\cite{Chen12}. In this latter protocol, the dissipation-free quantum trajectory can be parametrized as
\begin{equation}
\label{eq:STIRAPeigenstate}
| \psi_0(t) \rangle = \left( \begin{array} {c} \cos \gamma_0(t) \cos \beta_0(t) \\ - i \sin \gamma_0(t) \\ - \cos \gamma_0(t) \sin \beta_0(t) \end{array} \right)
\end{equation}
 Interestingly, the quantum state~\eqref{eq:STIRAPeigenstate} can be mapped onto a real-valued vector $\mathbf{S}_0(t)=   \cos \gamma_0(t) \sin \beta_0(t) \hat{\mathbf{x}} - \sin \gamma_0(t) \hat{\mathbf{y}} +\cos \gamma_0(t) \cos \beta_0(t) \hat{\mathbf{z}}$
   which behaves as an effective spin that obeys a precession equation: 
   \begin{equation}
  \label{eq:precessionbasicdissipation}
  \frac {d \mathbf{S}_0} {dt} = \gamma \: \mathbf{B}_0 \times \mathbf{S}_0.
  \end{equation}
where the effective magnetic field $ \mathbf{B}_0(t)=\frac 1 2 [\Omega_{p}^0(t) \hat{\mathbf{x}} + \Omega_{s}^0(t) \hat{\mathbf{z}}] $  is determined by the pump and Stokes fields.   
   
In the absence of dissipation, we introduce a reference trajectory, $\mathbf{S}_0(t)$, associated to a prescribed evolution of the angles $\beta_0(t)$ and $\gamma_0(t)$ that fulfills the required boundary conditions to ensure the transfer of the population from state 1 to state 3. The pump and Stokes fields $\Omega_{p 0}(t),\Omega_{s 0}(t)$ are subsequently inferred from the chosen trajectory (see Appendix A).  

 \subsubsection{The STIRAP solution in the presence of dissipation}

We now take into account dissipation. We assume that the intermediate level $| 2 \rangle$ suffers a finite damping,modelled by the anti-Hermitian Hamiltonian $\hat{H}_{\Gamma}= - i \hbar \Gamma_2  | 2 \rangle \langle 2 |$. The effective spin now obeys the differential equation
\begin{equation}
  \label{eq:precessionbasicdissipation}
  \frac {d \mathbf{S}} {dt} = \gamma \: \mathbf{B} \times \mathbf{S} -\overline{\overline{\Gamma}} \: \mathbf{S},
  \end{equation}
where the dissipation tensor is $\overline{\overline{\Gamma}}=\Gamma_2 \hat y \hat y$.  By superimposing to the original field $\mathbf{B}_0(t)$ the following geometric correction 
\begin{equation}
\label{eq:fieldcorrection}
\delta\mathbf{B}_0(t)= \gamma^{-1} \mathbf{S}_0(t) \times \overline{\overline{\Gamma}} \mathbf{S}_0(t),
\end{equation}
the effective spin  $\mathbf{S}$ follows the same trajectory as its dissipationless counterpart $\mathbf{S}_0$ despite the damping, or otherwise stated the renormalized state $| \widetilde{\psi}(t) \rangle = | \psi(t) \rangle / || | \psi(t) \rangle ||$ coincides with its dissipationless counterpart~\cite{Impens19}. The corresponding pulse corrections reads:
\begin{eqnarray}
\label{eq:pulsesCorrections}
\delta \Omega_p^0(t) & = & - {\Gamma_2} \: \sin 2 \gamma_0(t) \: \cos \beta_0(t), \nonumber \\
\delta \Omega_s^0(t) & = &  {\Gamma_2} \: \sin 2 \gamma_0(t) \: \sin \beta_0(t). 
\end{eqnarray}
The effective spin $ \mathbf{S}$ evolves in the magnetic field $\mathbf{B}(t)=\mathbf{B}_0(t)+ \delta \mathbf{B}_0(t)$. Interestingly, this approach restores the dissipation-free purity $p= p_{|3 \rangle} /(p_{| 1\rangle}+p_{| 2\rangle}+p_{| 3\rangle})  \simeq 99,8 \%$ of the final population in the target state. However, the quantum state norm $\mathcal{N}(t)=||| \psi(t) \rangle||$ may suffer a significant damping. The interest of the time-rescaling is also here to mitigate this latter effect. For the STIRAP problem, one readily derives the rescaled pulse fields:  $\Omega_{p,s}^{0 \: \Lambda}(t)  =  \dot{\Lambda}(t) \Omega_{p,s}^0(\Lambda(t))$ and $\delta \Omega_{p,s }^{0 \: \Lambda}(t)  =  \delta \Omega_{p,s}^0(\Lambda(t))$.
With the considered $\mathbf{S}_0$ trajectory, the population in the damped intermediate state $p(t)=| \langle 2 | \psi(t) \rangle |^2  = \sin^2(\gamma_0(t))$ reaches its maximum value at the half time $t=T/2$.  

\subsubsection{Comparing different time-scaled STIRAP solutions}

We propose hereafter two different time scalings that accelerate about this half time to reduce the norm decrease. First, we consider a polynomial scaling that fulfils this latter requirement $\Lambda_1(t)=T_1 P(t/T_1)$ with $P(x)=3x^2-2x^3$. Alternatively, we will consider a quasi-unitary time-scaling $\Lambda_2(t)$ (see Fig.~\ref{fig:STIRAP}b) associated to a uniform damping of the quantum state norm in the sense of~\eqref{requir1} and with respect to the geometric angle $\beta_0$. The scaling $\Lambda_2(t)$ is obtained by solving a differential equation analogous to~\eqref{eq:quasiunitarycondition}: 
\begin{equation}
\label{eq:constantdamping}
\dot{\Lambda}_2 =  \frac{\Gamma_2}{\dot{\beta}_0(\Lambda_2) c'}  \sin^2 \gamma_0(\Lambda_2) .
\end{equation}

Figure \ref{fig:STIRAP} compares the performance of the three protocols: the pulse sequence $\Omega_{p,s}(t)=\Omega_{p,s}^0(t)+\delta \Omega_{p,s}^0(t)$  and their time-scaled versions based on $\Lambda_1(t)$ and $\Lambda_2(t)$.

 For numerical applications,  we use the angular trajectories detailed in Appendix A parametrized with $\epsilon=0.05, \delta=\pi/4$ and for a damping rate equal to $\Gamma_2 T = 0.1$. First, we consider time-scalings $\Lambda_{1,2}(t)$ such that $\Lambda_{1,2}(T)=T$. This condition amounts to setting $T_1=T$ and $c' \simeq 4.94 \times 10^{-3}$. One obtains the 
respective quantum fidelities  $\mathcal{F}_0=0.954,$  $\mathcal{F}_{\Lambda_1, T}=0.966$ and $\mathcal{F}_{\Lambda_2, T}=0.982$ for respectively the initial protocol, for the polynomial scaling $\Lambda_1(t)$ and for the quasi-unitary scaling $\Lambda_2(t)$.  For the three protocols the final purity remains equal to the dissipation-free value $p \simeq 99,8 \%.$ The enhancement of the quantum fidelity results from a reduction of the norm $\mathcal{N}(t)$ damping.  

\subsubsection{Energetic cost and optimization}

Alternatively, one can choose the total duration $T_{1,2}$ of the time scalings $\Lambda_{1,2}(t)$ as to yield a protocol with the same energy as the original STIRAP protocol. The energy, taken as $E_{\Lambda_{k}}=  \int_0^{T_k} dt' ||\mathbf{B}_{\Lambda_{k}}(t')||^2 $, is proportional to the integrated Stokes and pump field intensities and inversely proportional to the total duration $T_k$. One finds the durations $T_1 \simeq 1.10 T$ and $T_2 \simeq 1.53 T,$ giving the quantum fidelities $\mathcal{F}_{\Lambda_1, T_1} \simeq 0.963$ and $\mathcal{F}_{\Lambda_2, T_2} \simeq 0.974$. At constant energy, the quasi-unitary time scaling $\Lambda_2(t)$ thus enables a reduction of the discrepancy $\tilde{\epsilon}=1-\mathcal{F}$ with the perfect transfer in the absence of dissipation by nearly $45 \%$ with respect to the original STIRAP protocol. This improvement at constant resources confirms the viability of the time-scaling technique.

\begin{figure}
\includegraphics[width= 0.45\textwidth]{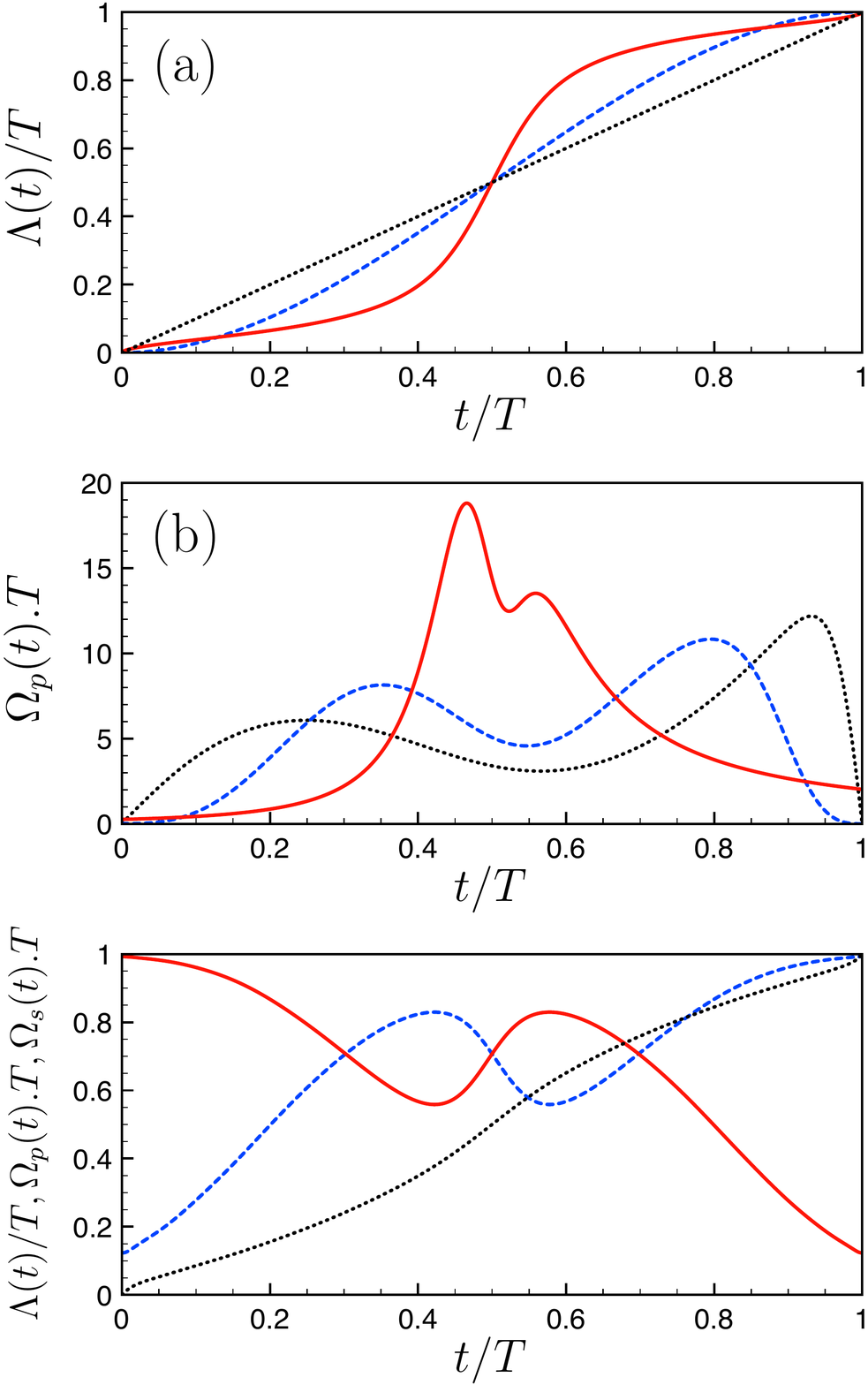}
    \caption{Application of the time-scaling to the STIRAP protocol:  (a): The scaling functions $\Lambda_1(t)/T$ (blue dashed line) and $\Lambda_2(t)/T$ (red solid line) as a function of the normalized time $t/T$ (with the choice $\Lambda_{1,2}(T)=T$). The black dotted line represents the trivial time scaling $\Lambda(t)=t$.  (b):  Rabi frequencies $\Omega_p$ (in units of $T^{-1}$) for the pump field for the dissipation-corrected STIRAP protocol~(\ref{eq:reverseegineeringpulses},\ref{eq:pulsesCorrections}) (black line), and its time-scaled versions for $\Lambda_1(t)$ (blue dashed line) and $\Lambda_2(t)$ (red dot-dashed line) as a function of the normalized time $t/T$. 
The original protocol is associated to the angular trajectories~\eqref{eq:angularfunctions} with the parameters $\epsilon=0.05, \delta=\pi/4$, and the dissipation rate is equal to $\Gamma_2 =0.1 /T$. (c) Time-scaled resource optimization of a STIRAP transfer:  Optimal time scaling $\Lambda(t)/T$ as a function of the normalized time $t/T$ (dotted black line) and the corresponding pump $\Omega_p$ (dashed blue line) and Stokes $\Omega_s$ (solid red line) Rabi frequencies $\Omega_{\Lambda s,p}(t)= \dot{\Lambda}(t) \Omega^0_{s,p}(t)$ renormalized by the constant value $\Omega_0=\sqrt{\Omega_{\Lambda p}^2(t)+{\Omega_{\Lambda s}^2(t)}}$.}  \label{fig:STIRAP}
\end{figure}

The previous formalism provides a strategy to determine a protocol that  minimizes the energy for the STIRAP transfer. First, we notice that even for strong dissipation rates such that $\Gamma_{\perp} T=1$ and  $\Gamma_{/ \! / \!} T=0.1,$ with the chosen angle $\epsilon=0.05$, the energy overhead associated to the correction remains small $\delta E_{\rm corr}/ E \simeq 0.6 \%.$ Regarding the optimization of the protocol through the time scaling, we can thus ignore the contribution associated with the correction and take $E[\Lambda,\dot \Lambda]=\hbar \int_0^T dt \dot{\Lambda}(t)^2 \left(\Omega_p^0(\Lambda(t))^2+\Omega_s^0(\Lambda(t))^2 \right)$, with the pump and Stokes fields~\eqref{eq:reverseegineeringpulses}. By minimizing this functional of $\Lambda$ and $\dot \Lambda$, we get the differential equation that $\Lambda$ obeys $\dot{\Lambda}(t) = c  \left (\dot{\beta}_0(\Lambda(t))^2 \cot^2 \gamma_0(\Lambda(t))+ \dot{\gamma}_0(\Lambda(t))^2 \right)^{-1/2}$. The constant $c$ is determined self-consistently by imposing the boundary value $\Lambda(T)$. The solution of the differential equation obeyed by $\Lambda$ imposes a constant norm for the effective field vector, $\Omega_\Lambda=\sqrt{\Omega_{\Lambda p}^2(t)+{\Omega_{\Lambda s}^2(t)}}=cte$. This optimal solution is reminiscent of the $\pi$-pulse optimal solution for the 2-level problem. Figure~\ref{fig:STIRAP}(c) represents this optimal time scaling $\Lambda(t)$ (for $\Lambda(T)=T$) as well as the pump and Stokes pulses obtained for such an optimization. The time-scaling accelerates when the effective magnetic field is minimal, as for instance at the initial and final times. With these optimal pulses, one obtains an energy $E_{\rm opt} \simeq 64.9 \hbar /T,$ which is is roughly $10 \%$ lower than the original pulse $E_{\rm 0}=  72.1 \hbar/T$ for the same purity $p=99.8$ \%.

\section{Time Scaling and Quantum Speed Limit for non-Hermitian Hamiltonians}
\label{sec:QSL}

The time optimality of a quantum state transfer is measured through the concept of QSL. Quantum systems evolve at a fraction of the QSL. This fraction constitutes a measure of the driving efficiency, and for an optimal driving it reaches unity. One can readily show that this driving efficiency is invariant under a time-scaling transform for a unitary evolution. Indeed, for closed quantum system the time-scaling changes equally the time and energy scales respectively related to the  quantum speed and to the QSL. In contrast, the dissipation is unaffected by the time-scaling transform and one would thus expect this invariance to break down in dissipative systems. In the following, we derive the expression of the QSL for dissipative systems modelled by a non-Hermitian Hamiltonian. Interestingly, with appropriate corrective terms in the quantum driving, the dissipation-free efficiency can be restored in a dissipative 3-level system. In this example, the driving efficiency remains invariant through time-scaling transforms even in the presence of dissipation.

\subsection{Quantum Speed Limit for non-Hermitian Hamiltonian}

The quantum speed limit (QSL) amounts to measuring the minimal time - associated to the maximal evolution velocity -  from a given initial state $|\tilde\psi(0) \rangle$  to a state orthogonal $|\tilde\psi(t) \rangle$ to the initial one (we denote $| \tilde{\psi}(t) \rangle = | \psi(t) \rangle / || | \psi(t) \rangle ||$.) 
 For a system evolving under the action of a time-independent Hamiltonian $\hat H_0$, it translates as an upper bound on the rate of variation of the angle $\cos \phi = |\langle \psi(0) | \psi(t) \rangle|$:
\begin{equation}
\frac{d \phi}{dt} \leq \frac{\Delta \hat H_0}{\hbar}.
\end{equation}
In appendix B, we show how Vaidman's derivation of the QSL in dissipationless system ~\cite{Vaidmann92} can be readily adapted to non-Hermitian time-dependent Hamiltonians $\hat H=\hat H_0-i\hat \Gamma$. The new bound reads
 \begin{equation}
\label{eq:QSLbound}
\dot \phi= \frac{d \phi}{dt}  \leq \frac {\sqrt{(\Delta \hat{H}_0)^2 + (\Delta \hat{\Gamma})^2 - i \langle [ \hat{H}_0, \hat{\Gamma}  ] \rangle}} {\hbar}.
\end{equation}
As a consistency check, we have performed numerical simulations in 2-level systems that confirm the validity of this upper bound.

\subsection{Quantum Speed Limit in a 2-level system}

Consider a 2-level quantum system where  $| e \rangle$ denotes the excited state and  $ | g \rangle$ the ground state. The QSL is saturated when $ \langle \widetilde{\psi}(t) | \widetilde{\psi}(0) \rangle \langle \widetilde{\psi}(0)  | \dot{\widetilde{\psi}}(t) \rangle$ is a real quantity where $| \dot{\widetilde{\psi}}(t) \rangle$ denotes the time derivative of the quantum state. For a generic parametrization of the state $| \widetilde{\psi}(\theta,\varphi) \rangle = a_t  | e \rangle +   b_t | g \rangle,$ and  the initial state $| \psi(0) \rangle= | e \rangle$, the saturation of the QSL bound is reached for $\dot{\varphi_e}=0$ with $a_t=|a_t|e^{i\varphi_e}$.  We define the control Hamiltonian as $\hat{H}_0(t)= \frac 1 2 \Omega_0 \left( | e \rangle \langle g | + | g \rangle \langle e |  \right)$ with a constant Rabi frequency $\Omega_0=\pi/ T$. We take into account the dissipation for both states thanks to the operator $\hat{\Gamma}= \Gamma_1 | e \rangle \langle e |  + \Gamma_2 | g \rangle \langle g |$. Under the Hamiltonian $\hat H=\hat H_0-i\hat \Gamma$, the quantum state $| \psi_t \rangle$ varies as a function of time with a coefficient $a_t$ real (i.e. $\varphi_e(t)=0$ at all times). As a result, the quantum speed is equal to the QSL with $(\Delta H_0)^2  (t)= \frac 1 4 \Omega_0^2(t)$, $(\Delta \hat{\Gamma})^2(t)= \Gamma_1^2 |a_t|^2+ \Gamma_2^2 |b_t|^2 - ( \Gamma_1 |a_t|^2+ \Gamma_2 |b_t|^2 )^2,$ and $i \langle [\hat{H}_0(t), \hat{\Gamma}] \rangle=\frac i 4 (\Gamma_1-\Gamma_2) \Omega_0(t) (a_t b_t^*- a_t^* b_t).$ 

To clarify how dissipation affects the quantum speed, we consider in the following two opposite cases: $\Gamma_e > \Gamma_g$ and $\Gamma_e < \Gamma_g$. In the first configuration, the faster decay of the excited state contributes to flip down the Bloch vector. One thus expects a quantum velocity faster than in the dissipation-free case. In the opposite sitaution ($\Gamma_e < \Gamma_g$), the ground state is less stable and one expects dissipation to slow down the quantum state transfer. Our expression for the QSL~\eqref{eq:defintionchi} captures this physics through the commutator $i \langle [\hat{H}_0(t), \hat{\Gamma}] \rangle:$ depending on the relative strength of the excited/ground state dissipation rates, this contribution increases or decreases the QSL. 

As an example, with the dissipation rates  $\Gamma_e T = 0.2$ and $\Gamma_g T = 0.01$, the quantum state evolves faster than in the dissipation-free system for $\Gamma_e > \Gamma_g$, and the $\pi$ pulse is achieved for $T^* \simeq 0.964 T$ while $T^* \simeq 1.0405 T$ when the values of the dissipation rates are exchanged. In both cases the damping seriously deteriorates the quality of the transfer and the final quantum fidelity. In these examples, the quantum speed reaches the QSL at all times. Such a saturation of the QSL persists after a time-scaling transform. More generally, we show below that the time-scaling transform can also preserve the ratio of the quantum speed to the QSL in a dissipative 3-dimensional system.


\subsection{Quantum Speed Limit in a dissipative STIRAP system}

We now consider the dissipative 3-level system of Section~\ref{subsec:TimeScalingSTIRAP}. We use the pulse sequence $\Omega_{p,s}(t)=\Omega_{p,s}^0(t) + \delta \Omega_{p,s}^0$corresponding to the sum of the dissipation-free pulses $\Omega_{p,s}^0(t)$~\eqref{eq:reverseegineeringpulses} and the associated dissipative corrections $\delta \Omega_{p,s}^0(t)$~\eqref{eq:pulsesCorrections}. Thanks to the pulse correction, the renormalized quantum state $| \widetilde{\psi}(t) \rangle$ follows exactly the dissipation-free trajectory, i.e.  $| \widetilde{\psi}(t) \rangle = | \psi_0(t) \rangle$ at all times. Thus, the angle $\phi(t)=\arccos \left( | \langle \widetilde{\psi}(t) | \widetilde{\psi}(0) \rangle |\right)$  fulfills $\phi(t)=\phi_0(t)$ at all times,  where $\phi_0(t)= \arccos \left( | \langle \psi_0(t) | \psi(0) \rangle |\right)$ is the angle associated to the dissipation-free trajectory. The effective quantum speed $\dot{\phi}(t)$ is thus given by the dissipation-free trajectory.
 
 In the corrected protocol, the non-Hermitian QSL $\chi(t)$ depends a priori on the dissipation-free control Hamiltonian $\hat{H}_0$, the geometric correction $\delta \hat{H}_0$ and the dissipation operator $\hat{\Gamma}$. We find $\chi(t)^2 =  (\Delta \hat{H}_0)^2  + \langle \{ \delta \hat{H}_0, \hat{H}_0  \} \rangle + (\Delta \delta \hat{H}_0)^2+ (\Delta \hat{\Gamma})^2 - i \langle [ \hat{H}_0, \hat{\Gamma} ] \rangle -  i \langle [\delta \hat{H}_0, \hat{\Gamma} ] \rangle.$ Remarkably, by using the explicit form of the geometric correction~\eqref{eq:pulsesCorrections}, the dissipative QSL $\chi(t)$ boils down to $\chi(t)= \Delta \hat{H}_0 (t)$. Thanks to the geometric correction, the non-Hermitian QSL~\eqref{eq:QSLbound} is exactly equal to the dissipation-free QSL~\eqref{eq:QSLNoDissipation} of the original protocol. The preservation of the dissipation-free quantum speed $\dot{\phi}_0(t)$ and QSL $\chi_0(t)$ despite dissipation comes at the price of an energy overhead associated to the extra term added to the Hamiltonian, $\delta \hat{H}_0$.  
 
As a corollary, when the time-scaling technique is applied, both the quantum speed and the dissipative QSL undergo similar transformations as $\dot{\phi}_{\Lambda}(t)= \dot{\Lambda}(t) \dot{\phi}_0(\Lambda(t))$ and $\chi_{\Lambda}(t) = \dot{\Lambda}(t) \chi_0(\Lambda(t)).$ That is to say, the time-scaling preserves the ratio $r_0(t)$ of the quantum speed to the QSL as  $r_{\Lambda}(t)=r_0(\Lambda(t))$ with $r_0(t)=  \hbar \dot{\phi}_0(t) / \dot{\chi}_0(t)$. Remarkably and thanks to the geometric correction, the dissipative dynamics keeps the same quantum speed and quantum speed limit as for the original dissipation-free protocol.

\section{Conclusion}

In conclusion, we have demonstrated the applicability and relevance of time scaling for quantum state transfer to optimize the resources and/or mitigate the effect of dissipation in non-Hermitian quantum systems. Actually, the scaling function provides an extra freedom in the system that can be used to minimize the energy or the norm reduction. The quantum speed limit has been here generalized to non-Hermitian Hamiltonians, and we have shown that time scaling does not affect the speed limit and thus the optimality when the system is driven at the quantum speed limit persists.

\begin{acknowledgments}
This work was supported by the Agence Nationale de
la Recherche research funding Grant No.~ANR-18-CE30-0013. F.I., F.M.D.A.., and F. A. P. thank CNPq, CAPES, and FAPERJ for financial support.
\end{acknowledgments}

\section*{APPENDIX A}
\label{eq:AppendixA}

We consider a polynomial form for the angular functions  $\gamma(t) = \sum_{j=0}^4 a_j t^j$  and $\beta(t) = \sum_{j=0}^3 b_j t^j $. They fulfill the boundary conditions of Protocol 2 of Ref.~\cite{Chen12}
  \begin{eqnarray}
  \label{eq:angularfunctions}
   & & \gamma_0(0) =\epsilon, \, \dot{\gamma}_0(0)=0, \,  \gamma_0(T/2)=\delta  \\
   & & \gamma_0(T)=\epsilon, \, \dot{\gamma}_0(T)=0   \nonumber \\
  & & \beta_0(0)=0, \, \beta_0(T)= \pi/2 \nonumber \\
& & \dot{\beta}_0(0) = 0, \, \dot{\beta}_0(T) = 0,  \, \gamma_0(T/2) = \delta \nonumber
  \end{eqnarray}
Such shortcut-to-adiabaticity solutions give rise to a trade-off between the amplitudes of Rabi frequencies and the transient population of the intermediate state $| 2 \rangle$ \cite{Chen12,FastSTIRAP21}.

The pump and Stokes fields $\Omega_{p 0}(t),\Omega_{s 0}(t)$ are subsequently inferred from the chosen trajectory:
\begin{eqnarray}
\label{eq:reverseegineeringpulses} 
\Omega_{p}^0(t)= 2 \left(\dot{\beta}_0(t) \frac {\sin \beta_0(t)} {\tan \gamma_0(t)} + \dot{\gamma}_0(t) \cos \beta_0(t) \right), \nonumber \\
\Omega_{s}^0(t)= 2 \left(\dot{\beta}_0(t) \frac {\cos \beta_0(t)} {\tan \gamma_0(t)} - \dot{\gamma}_0(t) \sin \beta_0(t) \right).
\end{eqnarray} 

A small angle initial angle $\epsilon$ is used, which yields an error $1-\mathcal{F}=O(\epsilon^2)$ for the protocol defined by Eq.~(\ref{eq:reverseegineeringpulses},\ref{eq:angularfunctions}) alone. For sake of simplicity, in our discussion on the quantum fidelity and of the quantum speed limit, we consider this protocol as such. However, a perfect transfer may be restored by adding an initial and a final stage to the STIRAP protocol~\eqref{eq:angularfunctions}, namely by using an initial/final small pulse of angle $\epsilon$ with the pump field $\Omega_p(t)$/Stokes field $\Omega_s(t)$ used separately. The full protocol would then corresponds to a sequence $| 1 \rangle \rightarrow \left( \cos \epsilon | 1 \rangle - i \sin \epsilon | 2 \rangle \right) \rightarrow \left(  - i \sin \epsilon | 2 \rangle  + \cos \epsilon | 3 \rangle \right) \rightarrow | 3 \rangle$. 

\section*{APPENDIX B: QUANTUM SPEED LIMIT FOR NON-HERMITIAN HAMILTONIAN}
\label{AQSL}

As a starting point, we remind that for any Hermitian operator $\hat A$ and any quantum state $| \psi \rangle$ \cite{Vaidmann92}
\begin{equation}
A | \psi \rangle = \langle A \rangle| \psi \rangle + \Delta A | \psi_\perp \rangle
 \end{equation}
where $| \psi_\perp \rangle$ is orthogonal to $| \psi \rangle$ and  $\Delta \hat{A}$  the variance of the operator $\hat{A}$.

Interestingly, this relation can be generalized to non-Hermitian Hamiltonian $\hat{H}= \hat{H}_0 - i \hat{\Gamma}$ (with the Hermitian operators
$ \hat{H}_0^{\dagger}= \hat{H}_0$ and $ \hat{\Gamma}^{\dagger}= \hat{\Gamma}$) on a given normalized quantum state $| \psi \rangle$. With the same notations as previously, and for any quantum state $| \psi \rangle$, we write
\begin{equation}
\label{eq:decomposition}
 \chi |\psi_{\perp} \rangle = \hat{H} |\psi \rangle - \langle \hat{H}  \rangle |\psi \rangle 
 \end{equation}
where $\chi$ is a positive real scalar (see below).  To get an explicit expression for the coefficient $\chi$, we write $\langle  \hat{H}^{\dagger} \hat{H}  \rangle = | \langle \hat{H}  \rangle|^2 + \chi^2$
and use 
$\hat{H}^{\dagger} \hat{H} =  \hat{H}_0^2 +  \hat{\Gamma}^2 - i [ \hat{H}_0, \hat{\Gamma} ].$
As a result, we find
\begin{equation}
\label{eq:defintionchi}
\chi = \left( (\Delta \hat{H}_0)^2 + (\Delta \hat{\Gamma})^2 - i \langle [ \hat{H}_0, \hat{\Gamma}  ] \rangle \right)^{1/2} \, .
\end{equation}
The anti-hermiticity of the commutator $ [ \hat{H}_0, \hat{\Gamma}  ]$ guarantees that the quantity $i \langle [ \hat{H}_0, \hat{\Gamma}  ] \rangle $ is real-valued.

In closed quantum systems, the usual definition of the quantum velocity rests on the fidelity with respect to the initial state $\mathcal{F}(t) = |\langle \psi(t) | \psi(0)\rangle|^2$ -- the quantum velocity is inversely proportional to the time for which this fidelity goes to zero.  By using the decomposition~\eqref{eq:decomposition} in the Schr\"odinger equation, one obtains the time derivative of the fidelity for non-unitary dynamics
\begin{eqnarray}
 \dot{\mathcal{F}}(t) & = &  -  2 \langle \hat{\Gamma} \rangle (t)  |\langle \psi(t) | \psi(0) \rangle|^2 \\
 & & -  \frac {2 \chi(t)} {\hbar}  \mbox{Re} \left[ i  \langle \psi(t) | \psi(0) \rangle  \langle \psi(0) | \psi_{\perp} (t) \rangle \right] \\
 & & = \dot{\mathcal{F}}_r + \dot{\mathcal{F}}_{\theta}\nonumber
 \end{eqnarray}
 The right hand side has two contributions with distinct physical interpretations. The first component $\dot{\mathcal{F}}_r= -  2 \langle \hat{\Gamma} \rangle (t)  |\langle \psi(t) | \psi(0) \rangle|^2 $  corresponds to a pure quantum state damping. 
 In contrast, the contribution $\dot{\mathcal{F}}_{\theta}$ accounts for a genuine rotation of the quantum state. $\dot{\mathcal{F}}_{\theta}$ is thus the only relevant contribution to the quantum velocity. 

We now propose a definition of the quantum velocity unaffected by the trivial quantum state damping. For this purpose, we introduce the renormalized quantum state  $| \widetilde{\psi}(t) \rangle =   | \psi(t) \rangle / \sqrt{ \langle \psi(t) | \psi(t) \rangle}$, and consider the  corresponding quantum fidelity $\widetilde{\mathcal{F}}(t) = |\langle \widetilde{\psi}(t) | \widetilde{\psi}(0)\rangle|^2$. By construction, only the relevant angular velocity $\dot{\mathcal{F}}_{\theta}$ contributes to the variation of this quantum fidelity,
i.e. $\dot{\widetilde{\mathcal{F}}}(t)=\dot{\widetilde{\mathcal{F}}}_{\theta}$. 

To determine an upper bound on $\dot{\mathcal{F}}_{\theta}$, we we apply the concepts introduced in Ref.~\cite{Vaidmann92}. The initial state can always be expanded over at most three orthogonal states as $| \psi(0) \rangle =
 \langle \widetilde{\psi}(t) | \widetilde{\psi}(0) \rangle  | \widetilde{\psi}(t) \rangle   +  \langle \widetilde{\psi}_{\perp}(t) | \widetilde{\psi} (0) \rangle   | \widetilde{\psi}_{\perp}(t) \rangle + \alpha | \widetilde{\psi}_{\perp \perp}(t) \rangle.$ This guarantees that $ |\langle \widetilde{\psi}(0) | \widetilde{\psi}_{\perp} (t) \rangle | \leq \sqrt{1-|\langle \widetilde{\psi}(0) | \widetilde{\psi} (t) \rangle |^2}$. As a result, 
$$
 |\dot{\widetilde{\mathcal{F}}}| = |\dot{\widetilde{\mathcal{F}}}_{\theta}| \leq \frac{2 \chi(t)}{\hbar} |\langle \widetilde{\psi}(0) | \widetilde{\psi} (t) \rangle|  \sqrt{1-|\langle \widetilde{\psi}(0) | \widetilde{\psi} (t) \rangle |^2}
$$
By introducing the usual definition $\cos \phi = \widetilde{\mathcal{F}}^{1/2}=|\langle \widetilde{\psi}(t) | \widetilde{\psi}(0)\rangle|$, we obtain the following upper bound for the quantum velocity
 \begin{equation}
\label{eq:QSLbound}
\dot{\phi}(t) \leq \frac {\sqrt{(\Delta \hat{H}_0)^2 + (\Delta \hat{\Gamma})^2 - i \langle [ \hat{H}_0, \hat{\Gamma}  ] \rangle}} {\hbar}.
\end{equation}
Other derivations of the QSL for dissipative systems, based on a matrix density formalism, can be found in Refs.~\cite{QSLOpen13a,QSLOpen13b}. Our non-unitary QSL has a similar form as the QSL obtained for closed quantum systems~\cite{Vaidmann92} 
 \begin{equation}
 \label{eq:QSLNoDissipation} 
 \dot{\phi}(t) \leq \frac {\Delta \hat{H}_0} {\hbar} 
 \end{equation}
  up to a replacement of the energy variance $\Delta \hat{H}_0$ by the quantity $\chi(t)$~\eqref{eq:defintionchi}. By the Ehrenfest's theorem, the variance $\Delta \hat{H}_0$ is time-independent for a unitary evolution in a constant Hamiltonian, leading to a constant QSL in this context. Nevertheless,  for the time-dependent and non-Hermitian Hamiltonians  considered here, the QSL $\chi(t)$ will generally vary with time.

  Our expression highlights the role of the dissipation operator in the evolution of the quantum state.  By construction the quantity $\chi^2=(\Delta \hat{H}_0)^2 + (\Delta \hat{\Gamma})^2 - i \langle [ \hat{H}_0, \hat{\Gamma}  ] \rangle$ is real-valued and positive, and is indeed bounded below by  $\chi(t)^2 \geq (\Delta H_0-\Delta \Gamma)^2 \geq 0$. This inequality shows that  a strictly positive quantum speed limit $\chi(t)>0$ exists for an eigenstate of the Hermitian Hamiltonian ($\Delta \hat{H}_0=0$) as long as dissipation has a strictly positive variance $\Delta \hat{\Gamma}>0.$

\end{document}